\documentclass[twocolumn, showpacs, prb ]{revtex4}

\usepackage{graphicx}

\usepackage{dcolumn}

\usepackage{bm}

\begin{document}

\title{Density dependence of valley polarization energy for composite fermions}

\date{\today}

\author{Medini\ Padmanabhan}

\author{T.\ Gokmen}

\author{M.\ Shayegan}

\affiliation{Department of Electrical Engineering, Princeton
University, Princeton, NJ 08544}

\begin{abstract}

In two-dimensional electron systems confined to wide AlAs quantum
wells, composite fermions around the filling factor $\nu = 3/2$ are
fully spin polarized but possess a valley degree of freedom. Here we
measure the  energy needed to completely valley polarize these
composite fermions as a function of electron density. Comparing our
results to the existing theory, we find overall good quantitative
agreement, but there is an unexpected trend: The measured composite
fermion valley polarization energy, normalized to the Coulomb
energy, decreases with decreasing density.

\end{abstract}


\maketitle

\section{Introduction}

When subjected to high perpendicular magnetic fields,
two-dimensional electron systems (2DESs) exhibit a wide variety of
exotic phenomena including the fractional quantum Hall effect
(FQHE). \cite{tsuiPRL82} The composite fermion (CF) theory
\cite{jainPRL89,halperinPRB93,CFbook} explains the FQHE of electrons
by mapping it to the integer QHE of CFs which are electron-magnetic
flux quasi-particles. Although absent in the simplest version of the
CF theory, the presence of discrete degrees of freedom, such as spin
and valley, ushers in a rich variety of phenomena. For many years,
understanding the spin polarization of the various FQHE states has
been a topic of great interest among experimentalists and theorists
alike. In the case of exotic states such as the one formed at Landau
level filling factor $\nu = 5/2$, the determination of
spin-polarization is valuable for deducing the nature of the ground
state as its possible non-Abelian statistics has promising
consequences for topological quantum computing. \cite{nayakRMP08}
For other states, e.g. those which form around $\nu =$ 1/2 and 3/2,
numerous transport,
\cite{eisensteinPRL89,eisensteinPRB90,engelPRB92,duPRL95,duPRB97}
optical, \cite{kukushkinPRL99,chughtaiPRB02} and nuclear spin
resonance and relaxation
\cite{dementyevPRL99,melintePRL00,freytagPRL02,tracyPRL07,liPRL09}
studies have aided the understanding of the role of spin in the CF
picture. \cite{parkPRL98,parkSSC01}

It was recently shown \cite{bishopPRL07} that CFs which form at $\nu
= 3/2$ in AlAs quantum wells possess a valley degree of freedom
which in principle is analogous to spin. In this study, we use
in-plane strain to tune the energy of the occupied valleys, and
measure the valley splitting energy needed to completely valley
polarize the CFs at and around $\nu = 3/2$. We find remarkably good
agreement between our results and the existing theory which was
developed to explain the spin-polarization of CFs in (single-valley)
GaAs. \cite{parkPRL98,parkSSC01} However, the polarization energy,
normalized to the Coulomb energy, is found to depend on the 2D
electron density ($n$), a feature not explained by the CF theory.

\section{Experiment details}

We report measurements on two samples (A and B) which are 12 and 15
nm-wide AlAs quantum wells grown using molecular beam epitaxy.
Details of sample growth are given in Ref. 21. A standard Hall bar
is fabricated using photolithography, GeAuNi alloy is used as
contact, and metallic front and back gates are deposited on the
sample which allow us to control $n$. Studies were done both in a
$^{3}$He cryostat and a dilution refrigerator at base temperatures
of about 0.3 K and $\simeq50$ mK, respectively and using standard
low-frequency lock-in techniques.

\section{System under study: A\lowercase{l}A\lowercase{s}}

\subsection{Effect of strain}

The band structure of bulk AlAs has three ellipsoidal conduction
band minima (valleys) at the X-points of the Brillouin zone. In
quantum wells wider than 5 nm, only two valleys with their major
axes lying in the 2D plane are occupied. \cite{shayeganPhysicaB06}
Their in-plane Fermi contours are anisotropic and are characterized
by transverse and longitudinal band effective masses,
\textit{m$_{t}$} = 0.205\textit{m$_{e}$} and \textit{m$_{l}$} =
1.05\textit{m$_{e}$}, where \textit{m$_{e}$} is the free electron
mass. The degeneracy between these two valleys can be broken by the
application of strain which we accomplish by gluing the sample to a
piezoelectric stack (piezo), as shown in Fig.\ 1(a). A voltage
$V_{P}$ applied to the piezo induces a strain
$\epsilon=\epsilon_{[100]}-\epsilon_{[010]}$ where
$\epsilon_{[100]}$ and $\epsilon_{[010]}$ denote strains along the
[100] and [010] crystal directions respectively.
\cite{shayeganPhysicaB06} This strain causes a transfer of electrons
from one valley to another. The resulting valley splitting energy is
given by $E_{v,e}=\epsilon E_{2}$ where $E_{2}$ is the deformation
potential which in AlAs has a band value of 5.8 eV.
\cite{shayeganPhysicaB06} Although the above picture of valley
occupation was first chalked out for the case of electrons,
\cite{gunawanPRL06} a similar approach for CFs was recently
demonstrated. \cite{bishopPRL07}

\subsection{Composite fermion picture around $\nu = 3/2$}
For the density range under study, the band parameters for AlAs
electrons are such that the Zeeman energy is larger than the
cyclotron energy. Since there are two valleys available for
occupation near $\epsilon = 0$, the first two electron Landau levels
(LLs) have the same spin. Since the CFs near $\nu = 3/2$ form in the
second electron LL, this CF system is effectively single-spin and
two-valley. This is in contrast to the energy level diagram for GaAs
electrons which involves only one valley; however the band
parameters in the GaAs system are such that the Zeeman energy is
small and the second electron LL is of the opposite spin. Thus, in
GaAs, the CFs formed around $\nu = 3/2$ form a single-valley,
two-spin system. In either case, the CF sea at $\nu = 3/2$ is formed
by hole-flux composite particles. \cite{CFbook} These CFs no longer
feel the externally applied perpendicular magnetic field, $B$.
Instead, they feel an effective magnetic field given by $B_{eff} = 3
(B - B_{3/2})$ where $B_{3/2}$ is the field at $\nu = 3/2$. The
various FQHE states formed around $\nu = 3/2$ are taken to be the
integer QHE states of these CFs. Each fractional electron filling
factor ($\nu$) has an integer CF counterpart ($p$). \cite{CFbook}

\section{Results and discussion}

In Fig. 1(c) we show magnetoresistance traces for sample A taken at
different values of $\epsilon$. At $\epsilon = 0$, the FQHE minima
at $\nu = 5/3$ and 7/5 are very weak or absent while the minimum at
$\nu = 8/5$ is strong. As we move away from $\epsilon = 0$ various
minima become weak and strong as a function of $\epsilon$. For
example, the traces shown in purple ($\epsilon = 0$) and blue
($\epsilon = \pm 0.64\times 10^{-4}$) show the weakest minima for
$\nu = 7/5$ while those in red ($\epsilon = \pm 0.40\times 10^{-4}$)
exhibit the weakest minima for $\nu = 8/5$. This behavior can be
qualitatively understood by following the simple CF energy fan
diagram shown in Fig. 1(b). At $\epsilon = 0$ each of the CF LLs is
doubly (valley) degenerate. This degeneracy is broken as we apply
strain and the two valleys separate in energy. There are specific
values of $\epsilon$ at which the LLs of CFs undergo an energetic
"coincidence" thereby causing the gap at the Fermi energy ($E_{F}$)
to vanish. For example, the $\nu = 7/5$ state (\textit{p} = 3) is
weak at $\epsilon = 0$ and undergoes one coincidence as $\epsilon$
increases before becoming completely valley polarized at large
$\epsilon$. In transport measurements, these coincidences show up as
the weakening of the FQHE minima. At high enough strains all the
states become fully valley polarized. Note here that as we apply
strain, both the electron and CF LLs split in energy. It is
important to realize that in the range of $\epsilon$ depicted in
Fig.\ 1, it is the CFs that become valley polarized; much larger
values of $\epsilon$ are needed to valley polarize the $electron$
LLs. \cite{bishopPRL07} This two-valley nature of the electron LLs
is critical for justifying the description of the CFs as being
formed in a "two-valley, single-spin" system.

\begin{figure}
\includegraphics[scale=1]{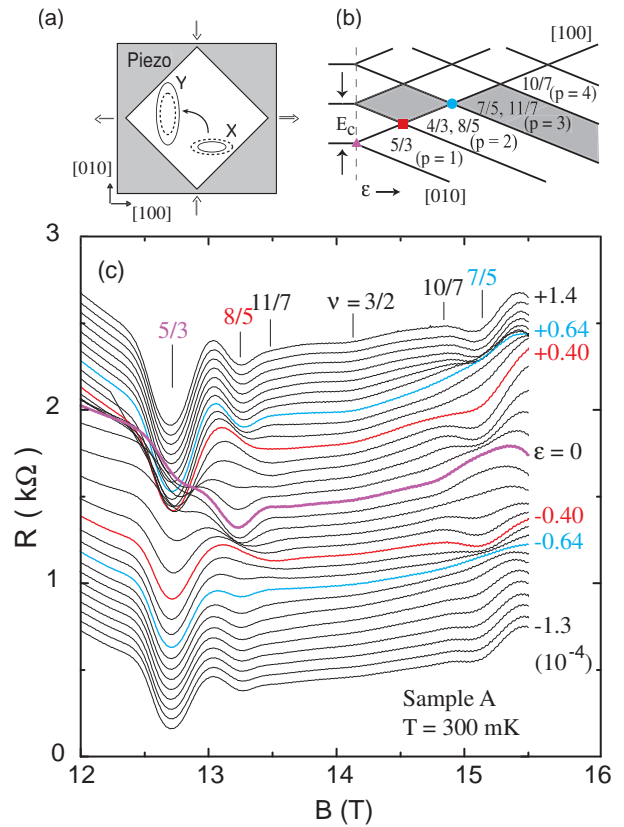}
\caption{(a) Experimental setup showing the sample glued to a piezo.
Schematic transfer of electrons from [100] to [010] valley in
response to strain is also shown. (b) Simple energy fan diagram in
the CF model for a two-valley, single-spin system. The [100] and
[010] valleys are degenerate in energy at $\epsilon = 0$ and the
degeneracy is lifted as strain is applied. (c) Magnetoresistance
traces taken at $T$ = 300mK, at different values of strain (shown on
the right in units of $10^{-4}$) for $n = 5.1 \times 10^{11}$
cm$^{-2}$. The traces are offset vertically for clarity.}
\end{figure}

\begin{figure}
\includegraphics[scale=1]{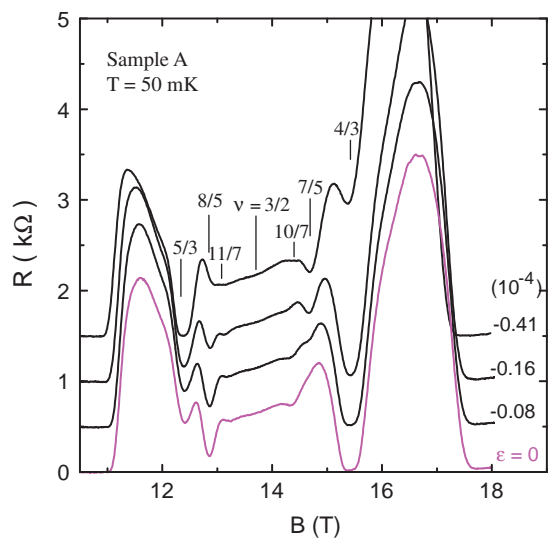}
\caption{ Magnetoresistance traces at $T$ = 50 mK for sample A taken
at different values of strain for $n = 5.0 \times 10^{11}$
cm$^{-2}$.}
\end{figure}

Note that the data in Fig.\ 1(c) were taken at $T$ = 300 mK. We
repeated these measurements at $T = 50$ mK in a second cooldown in a
different cryostat and the results are shown in Fig.\ 2. The
behavior is qualitatively the same, but we note that resistance
minima at some higher order fractions, for example $\nu = 11/7$ and
10/7 are better developed.

To demonstrate the response of the various minima to $\epsilon$, we
hold $B$ fixed at a particular $\nu$ and sweep $\epsilon$. In Figs.
3(a) and (b), we show results for sample A at $T$ = 300 mK and 50 mK
respectively. The peak positions are observed to be temperature
independent. Note the symmetry between the positive and negative
values of $\epsilon$. This is because the current in this sample is
flowing along the [110] crystal direction with respect to which the
[100] and [010] valleys are symmetric. In each trace of Fig.\ 3 the
phase and number of oscillations are consistent with the fan diagram
in Fig. 1(b). The high quality of this sample is evident from the
appearance of higher order fractions (up to $p = 4$). The
oscillations at the higher order fractions are particularly
interesting since the field sweeps show only weak evidence of their
existence. Similar measurements in sample B are shown in Fig.\ 4(a).

\begin{figure}[t]
\includegraphics[scale=1]{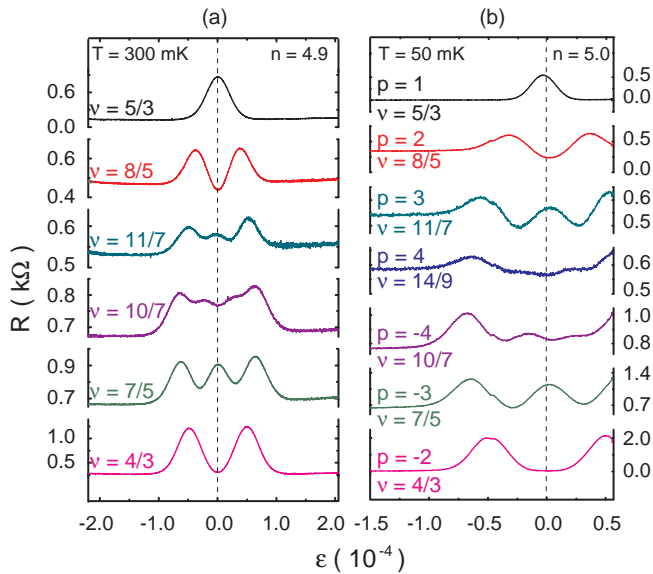}
\caption{ Piezoresistance traces taken for sample A at \textit{T} =
300 mK (panel (a)) and 50 mK (panel (b)) for different $\nu$. The
values of $n$ are given in units of $10^{11}$ cm$^{-2}$.}
\end{figure}

Our piezorsistance data allow us to determine the onset of CF valley
polarization at exactly $\nu = 3/2$ also. Notice the piezoresistance
trace taken at $\nu = 3/2$ in Fig. 4(a); for comparison, we also
show a trace at $B$ = 0 in Fig. 4(b). The $B$ = 0 trace can be
explained qualitatively in a simple model where the conductivities
of the two anisotropic valleys add in parallel, leading to an
increase in the resistance as the electrons are transferred from one
valley to the other. The piezoresistance can also stem from a loss
of screening as the electrons become valley polarized and lose their
valley degree of freedom; this is analogous to the
$magnetoresistance$ exhibited by 2DESs as the electrons become
$spin$ polarized in a parallel magnetic field. \cite{tutucPRL02} The
"kink" in the piezoresistance and the near-saturation of the
resistance at sufficiently large $\epsilon$ signal the full valley
polarization of the electrons, \cite{gunawanNP07} again, in analogy
with the magnetoresistance data. \cite{tutucPRL02} Now, as shown in
Fig. 4(a), we observe a qualitatively similar behavior at $\nu =
3/2$. The resistance at $\nu = 3/2$ exhibits a minimum when the two
valleys are balanced, increases as the valley degeneracy is broken
and saturates once the CFs are fully valley polarized. We take the
kink position, marked by an arrow in Fig. 4(a), to signal the
complete valley polarization of the CFs. \cite{footnote1} We note
that very recent experimental data \cite{liPRL09} for GaAs CFs at
$\nu = 1/2$ also show an enhancement, followed by a near-saturation,
of the resistance as the CFs become $spin$ polarized.

The energy needed to completely valley polarize the CFs,
$E_{v,pol}$, can be obtained directly from Figs.\ 3 and 4. For $\mid
p \mid =2$, 3 and 4, complete polarization is signalled by the
outer-most peaks in piezoresistances as these occur at the last
coincidence (see e.g., the red square and blue circle in Fig.\
1(b)). For the case of $\nu = 3/2$, the kink position in the
piezoresistance trace is taken as the point of full valley
polarization (shown in Fig.\ 4(a) with an arrow) with the upper and
lower excursions included in the error bars. In all cases,
$E_{v,pol} = \epsilon_{pol}E_{2}$ where $\epsilon_{pol}$ is the
measured threshold strain and $E_{2}$ = 5.8 eV.

\begin{figure}[t]
\includegraphics[scale=1]{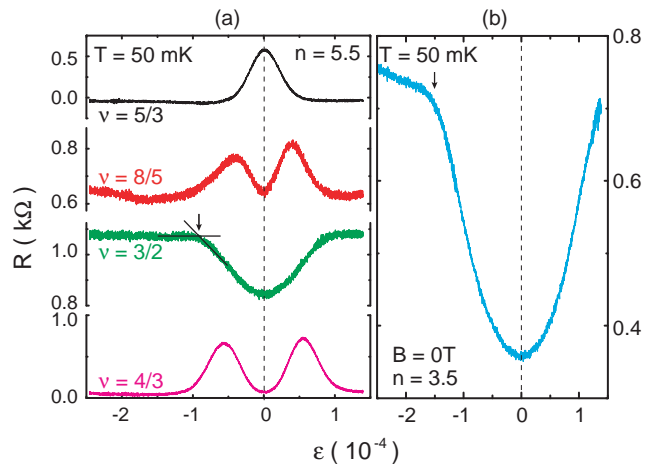}
\caption{(a) Piezoresistance traces taken for sample B at \textit{T}
= 50 mK for different $\nu$.  (b) Piezoresistance for sample B at $B
= 0$. The values of $n$ are given in units of $10^{11}$ cm$^{-2}$.}
\end{figure}

\begin{figure}
\includegraphics[scale=1]{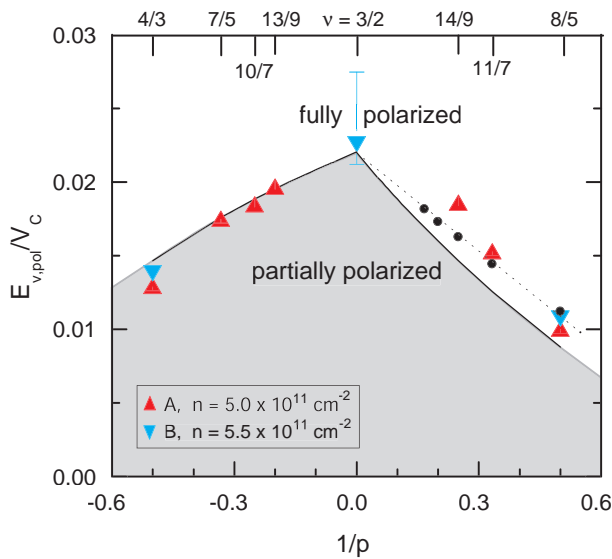}
\caption{Valley polarization energy in units of Coulomb energy as a
function of 1/$p$. Data are shown for both samples at comparable
$n$. Theoretical phase boundaries (originally worked out for
two-spin system) based on the microscopic CF theory (black circles)
and $m_{p}$ model (solid line) from Ref. 18 are also shown.}
\end{figure}

In Fig.\ 5 we plot $E_{v,pol}$ for various $\nu$ in units of the
Coulomb energy $V_{C} = e^{2}/4 \pi \kappa \epsilon_{0}\textit
l_{B}$ where $\kappa = 10$ is the dielectric constant of AlAs and
$\textit l_{B}=\sqrt{\hbar/eB}$ is the magnetic length. Data from
both samples (obtained from Figs. 3(b) and 4(a)) are shown at
comparable values of $n$. \cite{footnote2} Also shown are small
black circles (joined with a dotted line), taken from Ref. 18 which
is the theoretical calculation \cite{footnote3} using a microscopic
CF-wavefunction. Note that the theoretical phase boundary is
independent of $n$. This is not surprising since the CF-Hamiltonian
comprises entirely of the Coulomb interaction term. Hence all
relevant energy scales in the problem should scale as $V_{C}$. Since
the calculations are not done for negative values of $p$, we compare
our experimental data points to a simpler, one-parameter model which
is obtained in Ref. 18 as follows. The points obtained from the
microscopic CF theory for positive $p$ are extrapolated to $B_{eff}
= 0$ (1/$p$ = 0) to obtain an intercept of $E_{v,pol}/V_{C} =
0.022$. Since this value corresponds to the complete polarization of
the CF sea at $\nu = 3/2$, we have $E_{v,pol} = E_{F} = 2\pi
\hbar^{2} n_{CF}/m_{p}$ which gives $m_{p} = 0.47 \sqrt{B}$ in units
of $m_{e}$. \cite{footnote4} $m_{p}$ is defined to be the
\textit{polarization mass} of the CFs. \cite{parkPRL98} For a given
$p$, the condition of complete polarization can be written as $E_{v}
= (p-1)E_{c}$, where $E_{v}$ is the valley splitting of the CFs and
$E_{c} = \hbar e B_{eff}/ m_{p}$ is the CF cyclotron energy. Using
$B_{eff} = 3 (B - B_{3/2})$ and the fact that $\nu = 2 -
\frac{p}{2p\pm1}$, the phase boundary can be obtained in terms of
$p$ and $m_{p}$. The final result, $E_{v,pol}/V_{C} = 0.044
\frac{p-1}{2p\pm1}$, is shown as two solid curves separating the
partially- and fully-polarized CFs. Although there is no inherent
mass in the CF Hamiltonian, this simple one-parameter model is found
to be valuable in interpreting experimental data.
\cite{parkPRL98,parkSSC01}

There is overall good agreement between the experimental data points
and the theoretical phase boundary, given that there are no
adjustable parameters. Not only are the values very close to each
other, but also the asymmetry of the phase boundary about $B_{eff} =
0$ is reflected in the data. For example, $\nu = 4/3$ and 8/5 both
correspond to $\mid p \mid = 2$. The corresponding values of
$E_{v,pol}$, however, are theoretically expected to be different.
Consistent with this, our data shows that $E_{v,pol}$ for $\nu =
4/3$ is always larger than the value for $\nu = 8/5$.

However, this agreement might be fortuitous. First, both the
experiment and the theory have errors. Some causes of error in the
theoretical calculations for the ideal 2D system are examined in
Ref. 18. Our experiment carries an overall uncertainty in the
$y$-axis of up to 10$\%$ arising from our strain calibration.
Second, and more importantly, the experimental phase boundary for
the full valley polarization of CFs depends on $n$, in contrast to
the theoretical expectation.

To bring out the $n$-dependence of the phase boundary, we repeated
similar measurements for a range of $n$ in both samples and the
results are summarized in Fig. 6. The data for sample A were taken
at 300 mK (during the first cooldown) and data for sample B at 50
mK. The theoretical predictions based on the single $m_{p}$ model
are also shown as dotted lines. Our experimental results show that
there is a small, but measurable variation of the normalized
polarization energy as a function of $n$. Here we emphasize that,
for a given sample, the trend observed as a function of $n$ is
unaffected by the $\sim10\%$ error in our strain gauge calibration.

\begin{figure}[htcp]
\includegraphics[scale=1]{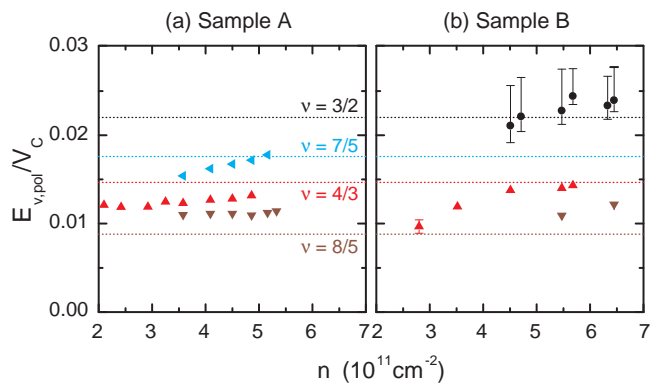}
\caption{(a) and (b) Valley polarization energy in units of Coulomb
energy plotted as a function of $n$ for samples A and B. The dotted
horizontal lines are theoretical predictions based on the $m_{p}$
model for the various $\nu$ as indicated. When not shown, error bars
are comparable to the symbol size.}
\end{figure}

The density dependence that emerges from our data demands a better
theoretical understanding. Note that residual interactions between
CFs cannot be blamed for the trend we observe. This is clear from
the $n$-independent nature of the phase boundary obtained from the
microscopic CF theory, which in principle includes effects of
interaction. \cite{parkPRL98} However, Ref. 18 deals with an ideal
2D electron system with no thickness, LL mixing or disorder. In Ref.
19, an attempt was made to address the role of finite layer
thickness. A simple way is to replace the bare Coulomb potential
$V_{C} = e^{2}/4 \pi \kappa \epsilon_{0}\textit l_{B}$ with an
effective potential $V_{C}' = e^{2}/4 \pi \kappa
\epsilon_{0}\sqrt{l_{B}^{2}+\lambda^{2}}$, where $\lambda$ is the
characteristic thickness of the electron wavefunction. In our
samples, where electrons are confined to square quantum wells, it is
reasonable to assume that $\lambda$ is constant or that it slightly
increases with increasing density. Either way, theory expects the
value of $E_{v,pol}/V_{C}$ to decrease as a function of increasing
$n$, opposite to the trend observed in our experiments. The effects
of LL mixing and disorder are unclear at the moment.

\begin{acknowledgments}
We thank the NSF for financial support. Part of this work was done
at the NHMFL, Tallahassee, which is also supported by the NSF. We
thank E. Palm, T. Murphy, G. Jones and J.H. Park for assistance and
J.K. Jain and K. Park for illuminating discussions.
\end{acknowledgments}

\break


\begin{references}
\bibitem{footnote2}
We have data from two other samples including the sample in Ref. 20.
The results are consistent with data shown in Fig.\ 5.
\bibitem{footnote3}
Although the theory deals with fractions around $\nu = 1/2$, it is
valid around $\nu = 3/2$ due to particle-hole symmetry. Also, the
theory deals with a two-spin system.
\bibitem{footnote4}
For GaAs, the corresponding expression is 0.60$\sqrt{B}$. The
difference in the prefactor is due to the difference in dielectric
constants of the host materials, GaAs and AlAs.
\end{references}
\end{document}